\def\etal{{\rm et al. }}
\def\mpc{{h^{-1} \rm Mpc}}
\def\mpcs{{h^{-1}_{70} \rm Mpc}}
\def\mpcsc{{h^{-1}_{75} \rm Mpc}}
\def\kpc{{h^{-1} \rm kpc}}
\def\kms{{\rm km\, s^{-1}}}
\newcommand\aap{{\em A}\&{\em A}}

\newcommand\aj{{\em AJ}}
\newcommand\apj{{\em ApJ}}

\newcommand\apjs{{\em ApJS}}
\newcommand\araa{{\em ARA}\&{\em A}}

\newcommand\mn{{\em MNRAS}}

\newcommand\nat{{\em Nature}}
\newcommand\nature{{\em Nature}}
\newcommand\pasp{{\em PASP}}
\newcommand\science{{\em Science}}

\documentclass[usenatbib]{mn2e}

\usepackage[dvips]{graphicx}
\usepackage{amssymb}
\begin{document}

\title{On the relation between Seyfert 2 accretion rate and environment at $z < 0.1$}

\author[Coldwell \etal]{Georgina V. Coldwell$^{1,2}$\thanks{E-mail:
gcoldwell@icate-conicet.gob.ar (GVC)}, Sebasti\'an Gurovich$^{3}$, 
 Jorge D\'iaz Tello$^{3}$, Ilona K. S\"ochting$^{4}$  \newauthor and Diego G. Lambas$^{3}$ \\
$^{1}$ ICATE-Conicet\\
$^{2}$ Universidad Nacional de San Juan, San Juan, Argentina\\
$^{3}$ IATE, Observatorio Astron\'omico, Universidad Nacional de C\'ordoba, 
Laprida 854, 5000, C\'ordoba, Argentina\\
$^{4}$ University of Oxford, Astrophysics, Denys Wilkinson Building, Keble 
Road, Oxford OX1 3RH, UK}

\date{\today}

\pagerange{\pageref{firstpage}--\pageref{lastpage}}

\maketitle

\label{firstpage}

\begin{abstract}

We analyse different properties of the small scale environment of Seyfert 2 for two
 samples selected according to the accretion rate parameter
 , $\cal R$, from the DR7-SDSS survey. We compare the results with
 two control samples of non-active
 galaxies that cover the same redshift range, luminosity,
 colours, morphology, age and stellar mass content.  Our study shows
 that both high and low accretion rate subsamples reside in bluer and
 lower density environments than the control samples. However, we find that this
 difference is at least two times stronger
 for the low accretion rate Seyferts.

In the vicinity of Seyfert 2, red galaxies have systematically 
lower values of stellar-mass as compared with corresponding control samples. The lower values 
of stellar mass for red neighbours is more significant at higher density environments and it is 
more evident for low accretion rate Seyfert. We also find that this effect is independent of the 
host's stellar mass. 

Our results are consistent with a scenario where AGN occurrence is higher in lower/medium 
density environments with a higher merger rate and a lack of a dense intergalactic medium 
(that can strip gas from these systems) that provide suitable conditions for the central black 
hole feeding. We find this particularly evident for the low accretion rate Seyferts that could 
compensate through the intergalactic medium the lack of gas of their hosts.

\end{abstract}


\section{Introduction}
 
The standard model that explains the widely accepted mechanism
responsible for the Active Galactic Nuclei (AGN) phenomenon is based
on the accretion of matter by massive black holes (BH) that lie at the
centres of galaxies \citep{LB69,Rees84}.  This effect manifests an
intense luminosity that can not be explained
by energetic processes generated by stars.  There is considerable evidence to suggest that most
massive galaxies harbour a central BH \citep{trem02,MH03},
however not all massive galaxies have AGN.  The underlying reason for
the observed occurrence or absence of AGN in massive galaxy hosts is a
question central to AGN study and has demanded considerable research
effort.

Although the precise details of how the gas cooling and migration
processes that trigger and influence AGN activity remain open to
debate, most theoretical models  \citep{OM89,Anto93,Urry95} suggest that all AGN 
have similar inner structure, and diversity mostly depends on the system 
inclination with respect to the observers line of sight. In this unified scheme, 
this relative geometry determines if broad or narrow line regions are
observed. If the torus is face-on, type 1 features are observed,
whereas if it is inclined broad-line (or type 1) features are
obscured by the dusty torus, and only type 2 features may be
observed. In this research we select type 2 AGN, and in
particular we focus on how Seyfert 2 accretion activity correlates with
galaxy properties in these environments.

The relative contribution of the processes that trigger and influence
AGN appear to be redshift dependent. In general, tidal torques
generated during galaxy interactions play a role in providing gas
inflows that feed central BHs \citep{sanders88}.  This could
occur during mergers \citep{roos81,roos85,gaskell85,HM95}, or galaxy
harassment \citep{LKM98}. Interestingly, the presence of multiple
BH/AGN in galaxies provides evidence that gas may even be
transported to the centers of galaxies during late stage
mergers \citep{BBR80}. Conversely, secular evolutionary processes
in galaxies that lead to disk instabilities also appear relevant since
they produce bars, rings, clumps and spiral structures that have been
proposed to play a significant role in transferring \citep{SBF90}
material towards galaxy centres, and there is evidence that the growth
of these structures may even be enhanced during low mass minor
mergers \citep{Purcell11}.  Moreover it seems reasonable to
consider that during galaxy and AGN evolution, more than one of the
aforementioned mechanisms may act simultaneously.

It seems clear that environmental factors affect the probability of
AGN occurrence and unraveling these factors will undoubtedly lead to
a more general understanding of AGN phenomenology. Considerable
progress has recently been made indicating that the galaxy number
density around low redshift AGN is similar to that around typical low
redshift galaxies found away from high density regions such as the
centers of galaxy clusters \citep{SBM95,coldwell06,Li06, SRR06}.
Moreover, although the density around AGN and typical galaxies in the low redshift 
universe are comparable, interestingly both quasars and AGN environments at lower
redshift are particularly populated by blue, disk-type and star-forming galaxies compared to
those in the vicinity of typical galaxies up to scales of $\sim 1 \mpc$ 
\citep{coldwell03,coldwell06}.

In low density environments the growth of disk instabilities occurs more
preferentially than in denser environments because abrupt dynamical
effects (ram pressure stripping, tidal forces, galaxy harassment,
major merging, etc) frequently found in galaxy cluster centres are
less efficient in removing and heating baryons. The decreased merger
efficiency in clusters due to the high relative velocities of cluster
galaxies is also a factor of AGN occurrence \citep{PB06}, since the
merger rate scales roughly as $\sigma_{v}^{-3}$ \citep{Mamon92,Makino97}.
Interestingly AGN have also been found between merging
clusters \citep{ilona02,ilona04} where the merger rate may also be
higher if progenitor relative velocities are smaller.

From a different perspective, \cite{padilla10} studied the correlation between the 
detected occurrence of AGN and the local and global environment, using a control 
sample of non-active galaxies restricting several parameters.
This allowed them to draw strong conclusions about the difference
between AGN host and non-active galaxy properties with respect to
environment. Their results indicate that for AGN host galaxies the
morphology-density relation is less significant in comparison with the
full SDSS galaxy sample and that AGN located closer to cluster galaxies tend
to be bluer than an early type control sample. This, may
indicate that AGN very close to high density regions need more
available gas to effectively feed the central black hole.

\cite{coldwell09} show that environmental
differences in the nearby universe are present in the blue and red AGN
population tails when compared to typical galaxy environments selected
to have similar host colour, redshift, luminosity, stellar mass ($M^{\ast}$), and
mean stellar age. The comparative results of \cite{coldwell09} reveal
the presence of a dichotomy: extremely red AGN hosts typically reside
in lower density environments than their non-active red control sample
that contain a lower fraction of blue star-forming galaxies. The blue
samples, both AGN and non-AGN, on the other hand have similar galaxy
number densities, specific star-formation rates and blue galaxy
fractions. Interestingly the morphology of both blue samples, AGN
hosts and non-active galaxies are significantly different such that a
higher fraction of early-type galaxies are found to host AGN as shown
in their Figure 5. Even though the morphologies of the red samples are
similar this is not reflected in their environments, therefore, the
notion of host-AGN co-evolution seems plausible, since different
processes may simultaneously affect the evolution of both the host and
AGN and these appear to be connected to environment. In this earlier
study LINERs and Seyfert 2 were both considered to be AGN type 2,
however in this study, we choose to exclude LINERs from the discussion
since recent studies have shown that significant numbers of LINER
emission is spatially extended and of stellar
origin \citep{GM09a,CapB11,YanB12}. 

On the nature of the typical environment around Seyfert 2 hosts, some analyses has been
completed. \cite{Taisa01} analysed a sample of 35 host galaxies with known AGN, and found a
higher percentage had close companions than their control sample
of non-Seyfert galaxies. \cite{dultin99} comparing the immediate
environments of Seyfert 1 and Seyfert 2 with those of two control
samples, find for the Seyfert 2 sample an excess of companions
within a projected distance $r_p< 100 \kpc$.  However,  as we show in our previous works, 
characteristics of the AGN
environment can be found within even larger scales of $\sim 1 \mpc$.
In this sense, \cite{Strand08} found that type II AGN are typically found in
denser environments than type I AGN on larger scales of the order of $\sim 2 \mpcs$.
Similarly, \cite{kolla12}, analysing SDSS-DR5 AGN, reported that within an environment of $1 \mpcsc$, 
Seyfert 1 galaxies have on average fewer companion galaxies than Seyfert 2 or HII galaxies.

To date, no large scale studies of the environment of a representative
sample of Seyfert 2 hosts (with control to match) has been made. 
The purpose of this study is to determine the characteristic nature of the
galaxy environment that favors Seyfert 2 occurrence and the dependence 
with the accretion rate of the BH. To do this we
discriminate our Seyfert 2 sample according to this parameter.

 The layout of this paper is as follows: In
Section 2 we briefly describe the Seyfert 2 classification scheme. 
Section 3 describes the Seyfert 2 subsamples host properties, their dependence 
with the accretion rate efficiency and the control samples selection.
Section 4 analyses the galaxy colours and number density
of galaxies surrounding our host Seyfert 2 subsamples. In
Section 5 we compare the surrounding galaxy mass distribution around
the subsamples. Finally in Section 6 we draw our main
conclusions and discuss our results. Throughout this paper, we have
assumed a $\Lambda$-dominated cosmology, with: $\Omega_{m} = 0.3$,
$\Omega_{\lambda} = 0.7$ and $H_0 = 100 \kms Mpc^{-1}$.

\section{Data and Sample Selection}

\begin{figure*}
\includegraphics[height=58mm,width=160mm]{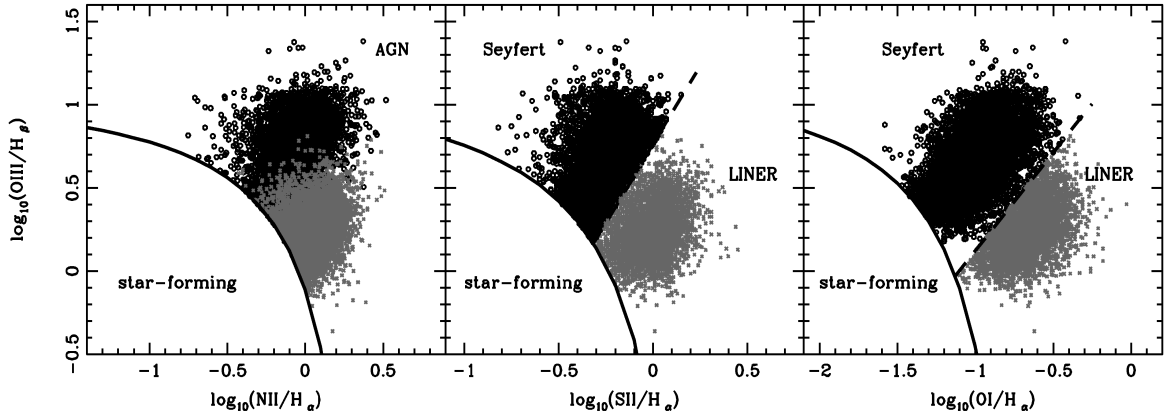}
\caption{The three BPT diagrams showing the selection criteria defined by Kewley et al. (2006) 
used to classify the emission-line galaxies as: Seyfert (black dots) and LINER (grey crosses) 
AGN. The solid lines separate the star-forming galaxies from the AGN and
the dashed lines represent the Seyfert-LINER demarcation.
}
\label{fig1}
\end{figure*}

The galaxy samples used in this analysis were taken from the spectroscopic 
Sloan Digital Sky Survey, Data Release 7 (SDSS-DR7: \cite{abazaja09}) which contains
$\sim 700000$ galaxies with measured spectra and photometry
in five bands, {\it u,g,r,i,z}. SDSS-DR7 main galaxy sample is essentially
 a magnitude \citep{petro76}  limited catalogue with \textit{$r_{lim}$}$ < 17.77$, 
with a median redshift of $0.1$ \citep{strauss02}.

For this study we use SDSS-DR7 physical galaxy properties derived and published  
by \cite{brinch04,tremonti04, blanton05} and available from MPA/JHU\footnote{http://www.mpa-garching.mpg.de/SDSS/DR7/}
and NYU\footnote{http://sdss.physics.nyu.edu/vagc/} that include: gas-phase metalicities, stellar masses, 
indicators of recent major star-bursts, current total and specific 
star-formation rates, emission-line fluxes, S\'ersic indices, etc. 

\subsection{Seyfert 2 Selection}
\label{sec:sel}

For the AGN selection we use the publicly available emission-line
fluxes. The method for emission-line measurement is detailed
in \cite{tremonti04}.  Additionally, we have corrected the
emission-line fluxes for optical reddening using the Balmer decrement
and the \cite{calzetti00} dust curve.  We assume an
$R_V=A_V/E(B-V)=3.1$ and an intrinsic Balmer decrement
$(H\alpha/H\beta)_{0}=3.1$ \citep{OM89}.  
Since the true uncertainties in the emission-line measurements were underestimated
the signal-to-noise ($S/N$) of every line was calculated with the emission-line flux 
errors adjusted according to the uncertainties suggested by the MPA/JHU 
catalogue\footnote{http://www.mpa-garching.mpg.de/SDSS/DR7/raw$\_$data.html}.

It is important to mention that although Seyfert 1 galaxies are not included in the main SDSS galaxy sample 
from which we derive the Seyfert 2 AGN, some Seyfert I contamination can be expected. In Seyfert 1 the 
continuum is affected by non-thermal emission difficulting the proper observation and study of their host galaxy. 
So, the rejection of type 1 AGN from the galaxy sample, automatically done by the SDSS spectral classification 
algorithm using Principal Component Analysis goes some way in purifying our sample. \cite{kauff03}, by inspection 
of the spectra of the most powerful AGN, estimates a contamination of approximately 8\% of type 1 AGN in the sample 
using a similar catalogue from an earlier release.

The redshift range of the main sample of emission-line galaxies was
restricted to: $0.04 < z < 0.1$. The lower limit prevents that small
fixed-size apertures affects galaxy properties as derived by the fiber
spectra, and the upper limit corresponds to the luminosity
completeness limit for the SDSS sample. Furthermore, we conservatively
include only galaxies with signal-to-noise ratio (S/N) $> 2$ for all
the lines used in the three diagnostic diagrams to discriminate
Seyfert 2 from star-forming and LINER galaxies. 
We note that the
S/N cut was selected even though the adjusted uncertainties
almost duplicate the original errors and moderately reduce
our available sample, however in being more
conservative we can be more confident of LINER, Seyfert 2 and starburst
host membership, their respective properties as derived and conclusions drawn. 
Thus,
from this refined sample, we separate Seyfert 2, LINER and
star-forming using the three standard \cite[][: BPT]{BPT81} line-ratio
diagrams. AGN/starburst separation, as suggested
by \cite{Kewley01,Kewley06}, depends on the relative source location within BPT
diagnostic diagrams, following Eqn: \ref{eqn:bpt1} to \ref{eqn:bpt3}

\begin{equation}
\log([\rm OIII]/\rm H\beta) > 0.61/(\log(\rm [NII/H\alpha])-0.47)+1.19,
\label{eqn:bpt1}
\end{equation}

\begin{equation}
\log([\rm OIII]/\rm H\beta) > 0.72/(\log(\rm [SII/H\alpha])-0.32)+1.30,
\label{eqn:bpt2}
\end{equation}

\begin{equation}
\log([\rm OIII]/\rm H\beta) > 0.73/(\log(\rm [OI/H\alpha])+0.59)+1.33,
\label{eqn:bpt3}
\end{equation}

where Seyfert 2 are located above the Seyfert-LINER line on the 

\begin{equation}
\label{eqn:bpt4}
\log([\rm OIII]/\rm H\beta) > 1.89\log(\rm [SII/H\alpha])+0.76,
\end{equation}

\begin{equation}
\label{eqn:bpt5}
\log([\rm OIII]/\rm H\beta) > 1.18\log(\rm [OI/H\alpha]+1.30.
\end{equation}

We emphasize that LINERs are not considered bonafide AGN and hence
removed using Eqns \ref{eqn:bpt4} and \ref{eqn:bpt5} and the environmental analysis 
of the LINERS subsample  will be left for a future
study. For the sake of internal consistency we also exclude ambiguous
galaxies classified as one type of object in one BPT and another in
the remaining two diagrams. Bearing this in mind, we obtain a sample
of 4035 Seyfert 2. The discriminated samples and the selection
criteria are shown in the three BPT diagrams of Fig. 1.

In addition, to test for contamination of Seyfert 1 on our selected sample 
we study the Full Width at Half Maxima ($FWHM$) distribution of the $H\alpha$
emission lines. \cite{Ho05} suggest that a $H\alpha$ $FWHM=1200 \kms$ provides 
suitable separation between narrow-line (Seyfert 2) and broad-line (Seyfert 1) AGN. 
Taking into account this criteria only a 5\% of the selected AGN have values of 
$H\alpha$ $FWHM > 1200 \kms$. These AGN were removed from our sample to minimize the Seyfert 1 
contamination.

\section{Dependence of the host properties and the BH accretion efficiency}

The BH growth rate (in mass) depends on AGN activity. In the local
universe BHs with the highest growth rates are found at the centres of
late-type galaxies that are presently undergoing important
internal/external accretion. In comparison more massive BHs have
growth rates that are $\sim 10-1000$ times less efficient since they
have already accreted a significant fraction of their masses at higher
redshifts during major mergers \citep{goulding10}. We set out to
extend on previous results and explore other factors that might
influence BH growth rates in the local universe. In particular with
our SDSS DR7 sample we examine if host galaxy properties and
associated environmental parameters might correlate with BH growth rates at
different scales.

To estimate the BH accretion rate ($\mathcal R$) we use the standard correlation between
the  BH mass ($M_{BH}$) and the bulge velocity dispersion ($\sigma_*$) \citep{trem02} 
of Eqn \ref{eq:bh}.

\begin{equation}
\label{eq:bh}
\log{M_{\rm BH}} = 8.13 + 4.02\, \log{(\sigma_*/ 200\kms)}
\end{equation}

The velocity dispersion values were taken directly from the MPA/JHU catalogues\footnote{http://www.mpa-garching.mpg.de/SDSS/DR7/raw$\_$data.html}. 
We note that the analysis is restricted to Seyfert 2 hosts with $\sigma_*$ $>$ 70 km $s^{-1}$
given that the instrumental resolution of the SDSS spectra is $\sigma_*$ $\approx$ 60 to 70 km $s^{-1}$. 
This $\sigma_*$ restriction places a lower limit of black hole mass of $10^{6.3} M_\odot$.

Some of the dispersion in the $M_{\rm BH} - \sigma_*$ relation may result from different galaxy 
morphological types as suggested by \cite{GadoK09,Grene10}. 
The intrinsic scatter in $M_{\rm BH} - \sigma_*$ relation of early-type galaxies is smaller than 
that for late-type galaxies \citep{Graham08,Gulte09}. Previous results 
\citep{kauff03,heck04} found that the AGN, of all luminosities, reside mainly in galaxies with prominent 
bulges such as those of early type galaxies. So, we expect a negligible impact of this estimation in our 
analysis.

The AGN $\mathcal R$  parameter can be determined following \cite{heck04}, using the dust-corrected [OIII]
 $\lambda5700$ luminosity (see Sec. \ref{sec:sel}) as a tracer of the
 AGN activity \citep[see for instance,][]{kauff03, heck05}.

The $L[OIII]$ values due to AGN activity could be overestimated because of contamination by star-forming regions, 
in particular for low luminosity AGN. \cite{heck04} compute the average AGN contribution to $L[OIII]$ values for
a sample of composite galaxies, located above the \cite{kauff03} line but below the \cite{Kewley01} line in the 
BPT diagram, and then compared these values to those for AGN-dominated galaxies, above the \cite{Kewley01} line 
with SDSS data. The results of their work indicate that the $L[OIII]$ values of composite galaxies from AGN 
activity contribute between 50\% to 90\% of the total $L[OIII]$, while for AGN-dominated galaxies more than 
90\% of the $L[OIII]$ is due to the AGN. Their results agree with that reported by \cite{KauHe09}.

For the Seyfert 2 selection we have used the
conservative boundary of \cite{Kewley01} and, in
addition, have excluded LINER galaxies from the
sample by minimizing any contamination by low luminosity AGNs and/or composite galaxies. Therefore, we do not 
expect any significant contribution
by star-forming regions on our $L[OIII]$ values for our Seyfert 2 sample.

The AGN $\mathcal R$  parameter is estimated by Eqn \ref{eq:abh}.

\begin{equation}
\label{eq:abh}
\mathcal{R}=\log(L_{\rm{[OIII]}}/M_{\rm{BH}})
\end{equation}

\begin{figure}
\includegraphics[width=90mm,height=80mm ]{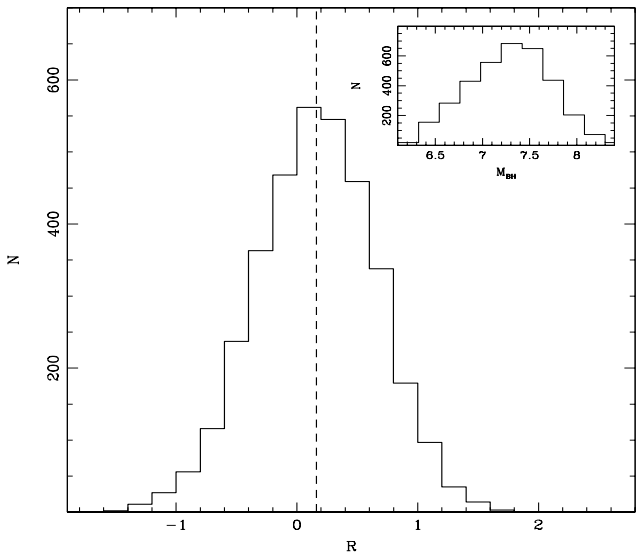}
\caption{Distribution of the accretion rate parameter for the Seyfert 2 sample. The dashed line indicates the average of the
distribution. Inner box: BH mass distribution.
}
\label{hR}
\end{figure}

In Fig. \ref{hR} the distribution
of the $M_{\rm BH}$ (inset box) and $\mathcal{R}$ values for our Seyfert 2 is shown.
With the aim to investigate the dependence of the accretion rate efficiency of Seyfert 2 with 
the environment we divide this sample into high
(Seyfert-HR) and low (Seyfert-LR) $\mathcal R$ systems, respectively,
using the mean value of the distribution (dashed line Fig. 2) as our
limit producing a final sample of 1784 Seyfert-HR and 1731 Seyfert-LR.

\begin{figure*}
\includegraphics[width=180mm,height=95mm,]{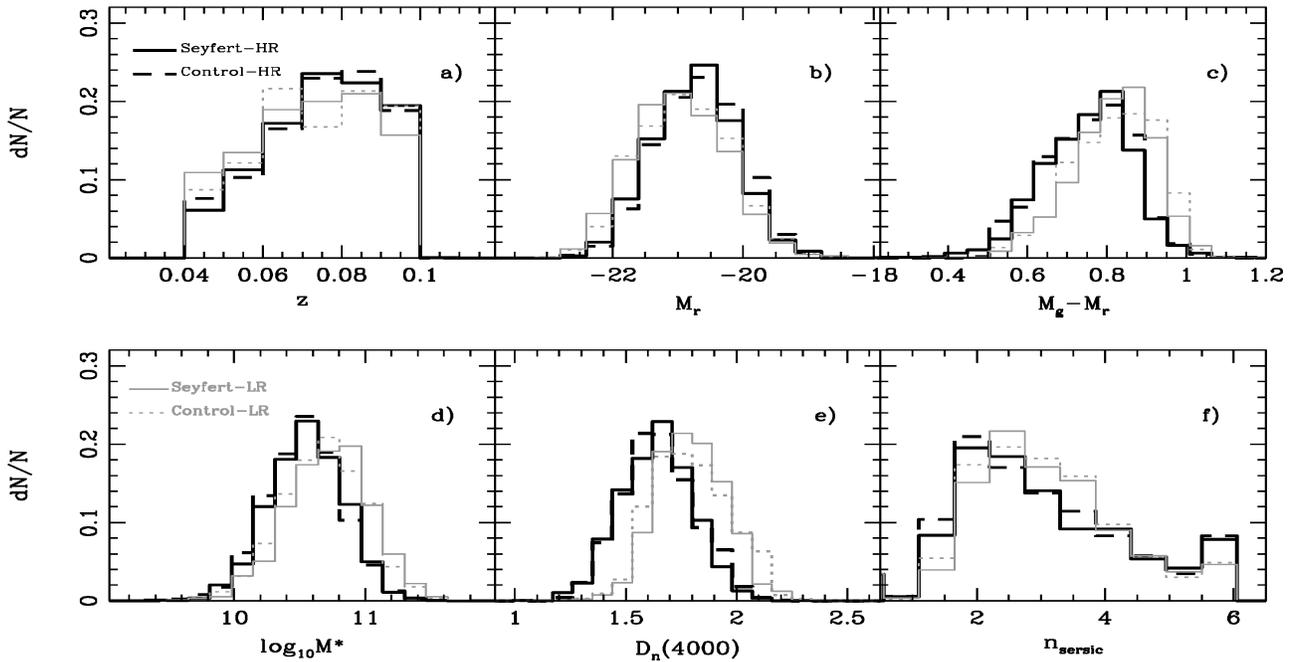}
\caption{Normalized distributions of matched system properties for: Seyfert-HR (black solid lines),
Seyfert-LR (grey solid lines) and their respective control samples, Control-HR galaxies (black dashed lines)
and Control-LR galaxies (grey dotted lines).}
\label{fig:dist}
\end{figure*}

Figure \ref{fig:dist} shows the distributions of host properties of
our divided samples, in the panels labeled for: $a)$ Redshift, $b)$
Extinction and K-corrected \citep{blanton03} absolute r-band magnitude
(Petrosian), $c)$ colour ($M_g-M_r$), $d)$
stellar mass, $M^{\ast}$, previously determined by
\cite{kauff03b} \footnote{The method relies on spectral indicators relating to the stellar age 
and the fraction of stars formed in recent bursts.}, $e)$ Break index,
$\rm D_n(4000)$ defined in \cite{kauff02} as a proxy for the mean age of
the stellar population \footnote{We note that the majority of star formation
takes place preferentially in galaxies with low $\rm D_n(4000)$
values.}, and $f)$ S\'{e}rsic index \citep[][]{sersic63}. 

The S\'ersic Law, a generalization of the de Vaucouleurs and
Freeman laws, relates the surface brightness variation with distance
from the galaxy centre. The S\'{e}rsic law has the form $\ln I(R)= \ln
I_{0} -kR^{1/n}$, where $I_{0}$ is the intensity at $R = 0$. The
S\'{e}rsic index $n$, controls the degree of curvature of the
profile. The best-fit value of $n$ correlates with galaxy size and
luminosity such that bigger and brighter galaxies tend to be fit with
larger values of $n$ \citep{caon93, YC94}. Most galaxies are fit by
S\'{e}rsic profiles with indices in the range $0.5 < n < 10$. Setting
$n = 4$ gives the de Vaucouleurs profile which is a good description
of giant elliptical galaxies. Setting $n = 1$ gives the Freeman
exponential profile which is a good description of the light
distribution of both disk galaxies and dwarf elliptical galaxies.

In Fig. \ref{fig:dist} the Seyfert-HR and Seyfert-LR appear to belong
to different galaxy populations, as reflected by the host galaxy
properties. Host galaxies of Seyfert-HR tend to be bluer and younger
than those of Seyfert-LR. This indicates that the $\mathcal R$ is
somewhat dependent on galaxy age as seen in the $\rm D_n(4000)$ and
colour distributions.  The normalized numbers of Seyfert-HR and
Seyfert-LR also change, more lightly, with host luminosity and $M^{\ast}$ such that at
low luminosity and $M^{\ast}$ there is a larger relative number of
Seyfert-HR compared to Seyfert-LR. Conversely at high luminosity and
$M^{\ast}$ this tendency is reversed. Similarly, the relative fraction
of disk host ($n<~2$) with Seyfert-HR is higher than those containing
Seyfert-LR for the same $n$. These trends are consistent with the
general idea that BH in disk galaxies are the fastest growing in the
local universe \citep{goulding10}. We wish to note that both Seyfert
samples have similar redshift distribution so we can be more confident
of our results.  These results imply that typically both the host and
BH properties appear to co-evolve. When the host gas content is evolved
or depleted the $\mathcal R$ should also diminish. However cases may
also exist in which either an appropriate environment may favor
additional activity or an external event that may reactivates the
accretion processes \citep{coldwell09}.

\subsection{Galaxy control samples for Seyfert 2 hosts}

There are noticeable differences (as shown) between the host properties
of Seyfert-HR and Seyfert-LR and we might expect that characteristics
of their environment may reflect these differences. As has been
shown, environments of non-active galaxies may differ from
those of AGN \citep{coldwell03,coldwell06}.  Hence, an
appropriate study of the neighbouring galaxy properties to Seyfert-HR
and Seyfert-LR is much needed. 

In a series of papers \cite{coldwell03,coldwell06} and \cite{coldwell09}, 
use control samples to understand the behavior
of AGN with respect to non-active galaxies and explore their
relations with the large and small-scale environment. In almost every
case, barring when the AGN hosts are extremely blue, they find that
the AGN neighbourhood is populated by bluer and more star-forming
galaxies than for the control samples.  However, when the AGN hosts
are extremely blue, the non-active galaxy counterparts are similarly
blue but with different morphology. Nevertheless the environments of
both AGN and control samples are quite comparable in terms of number
density and general surrounding galaxy properties. These previous results suggest
that host morphology and environment, both play a role at triggering or
sustaining nuclear activity.

The selection of an appropriate control sample is
substantially important to obtain conclusions by direct comparison
with galaxies without detected nuclear activity.
We compare environmental properties of the Seyfert-HR and Seyfert-LR
with the environments of two corresponding control samples of
non-active galaxies. The control sample is constructed with galaxy
properties matching those of both Seyfert 2 samples, respectively in
redshift, luminosity, colour, stellar mass, mean stellar age, and
morphology. The Seyfert-LR and Seyfert-HR control sample properties
are shown in Fig. \ref{fig:dist} (as dotted and dashed lines,
respectively).  These restrictions ensure that any environmental
difference is most likely due to differences in nuclear
activity. With these restrictions we have 3246 galaxies in
the Control-HR sample, and 3165 galaxies corresponding to the Control-LR.

\section{Properties of the surrounding galaxies}

Galaxy colours provide an indirect constraint on the evolutionary
history of galaxies since many galaxy parameters for example
morphology, age, [Fe/H], environmental density all, affect galaxy
colours. In high density regions, such as galaxy clusters, the large
fraction of red galaxies indicates galaxies of an older stellar
population with low star-formation. Whereas galaxies in poor groups
and in the field in general have bluer colours with enhanced
star-formation. This motivates us, as in previous studies, to use $\rm
M_g-M_r$ colours to study galaxy properties around AGN and control
samples, as well as to constrain properties of their environments.

In order to quantify any excess of blue galaxies in the AGN
environments, we calculate the fraction of galaxies bluer than $\rm
M_g-M_r = 0.75$ as a function of the projected distance to the sample
centres\footnote{This fiducial colour is chosen since it approximately corresponds to the mean
colour for the spectroscopic survey at $z < 0.1$}.  The selected
tracer galaxies have projected distances $r_p < 2 \mpc$ and radial
velocity differences $\Delta V < 500 \kms$ , relative to the target
systems.

\begin{figure}
\includegraphics[width=90mm,height=110mm]{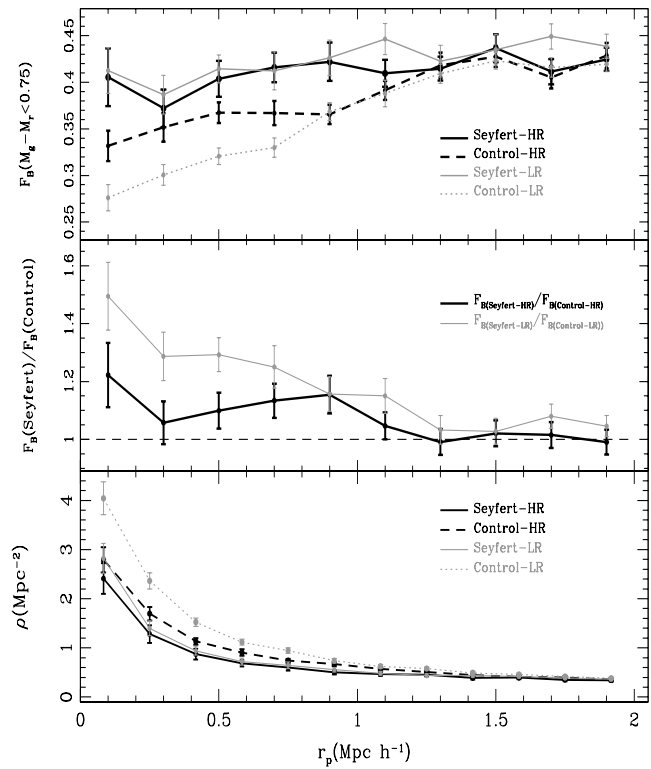}
\caption{Top: Fraction, $F_B$, of $\rm M_g-M_r < 0.75$ blue galaxies around the  
targets as a function of the projected distance $r_p$. 
Middle: Blue galaxy ratio between Seyfert and the control samples.
Bottom: Number density of galaxies around the different target samples.}
\label{fig:frac}
\end{figure}

The fraction of blue neighbouring galaxies is shown in the top panel
of Fig. \ref{fig:frac}.  
The error bars in this figure and in all figures were calculated by using bootstrap 
error resampling technique \citep{barrow84}. Around 40 $\%$ of Seyfert 2 neighbours are
bluer than the fiducial mean irrespective of $\mathcal R$. However, a
significant difference exists between the Seyfert 2 neighbourhoods and
the respective neighbourhoods centered on their non-active
counterparts whereby Seyfert 2 have typically bluer environments.
 This is evident up to $r_p \sim 1 \mpc$ and 
becomes more notorious at scales smaller than $r_p \sim 300 \kpc$. 

This effect is
more clearly observed in the middle panel showing the ratios between
the fraction of blue galaxies around Seyfert-HR and Seyfert-LR with their
respective control samples. The rate between Seyfert-LR and Control-LR
samples is approximately 30\% higher as compared to the ratio between
Seyfert-HR and Control-HR within $r_p < 300 \kpc$ and this effect
extends up to $r_p \sim 700 \kpc$.  We note that this effect is
notably diluted if we use a radial velocity difference criterion of
$\Delta V < 1000 \kms$.  Thus, although both Seyfert 2 host galaxies
and their matched non-active control sample have similar physical
properties (morphology, luminosity, stellar mass and age), their
environments do not reflect this fact. Therefore it seems plausible
that within the approximate scale size of a galaxy group, co-evolution
of the host galaxy and associated AGN activity is likely to depend on
environment.

Thinking in terms of the available gas reservoir to the AGN host
galaxy, if we assume the Seyfert-LR host is sufficiently evolved and
massive, and the amount of gas is insufficient to enhance BH activity,
one way to sustain gas accretion would be via gas-dynamical processes
that are more frequent in environments populated by a larger numbers
of blue, disk-type, star-forming galaxies \citep{coldwell09}. On the other
hand if we assume the Seyfert-HR host is a less massive evolving
galaxy, to sustain its enhanced AGN activity a smaller amount of gas
would be required and this is reflected in the smaller difference
between blue fractions of Seyfert-HR neighbouring galaxies and
Control-HR neighbouring galaxies. Furthermore, the increasing of the
blue neighboring fraction, from the centres up to 
$r_p \sim 1 \mpc$ of the
control samples (more evident in the Control-LR) would suggest that
the standard morphology-age-density relation is followed, since as is
well known, mean colours of neighbouring galaxies correlate to
differences in the environmental density of
galaxies \citep{Dress80,dominguez01}. However this relation does not
appear to be followed in the neighbourhood of Seyfert 2 hosts, as was
studied by \cite{coldwell06,coldwell09}.

We find significant differences in $M_g-M_r$ at scales smaller than
$r_p \sim 1 \mpc$ that relates to AGN activity. Since we wish
to examine the morphology-age-density relation further we calculate the galaxy number
density around both Seyfert and Control samples within cylinders of
projected radius $r_p$ and $\Delta V < 500 \,\kms$ in depth. The total
local number density of galaxies around every target is shown in the
bottom panel of Fig. \ref{fig:frac}.  The observed trends indicate
that both control samples are located in higher density environments
than their active counterparts at the smallest scales probed, a trend
that is significantly more pronounced between the Seyfert-LR and
Control-LR.  This possibly occurs because AGN tend to avoid regions of
higher density since in denser environments the AGN occurrence and
accretion rate is suppressed. This is in agreement with results
determined by \cite{PB06} and by \cite{coldwell09}. Instead, in low
density environments, especially those with an excess of blue
galaxies, a higher galaxy merger rate could play an important role in
the AGN occurrence. 

\section{Galaxy densities and stellar mass profiles in AGN neighbourhoods}

\begin{figure}
\includegraphics[width=90mm]{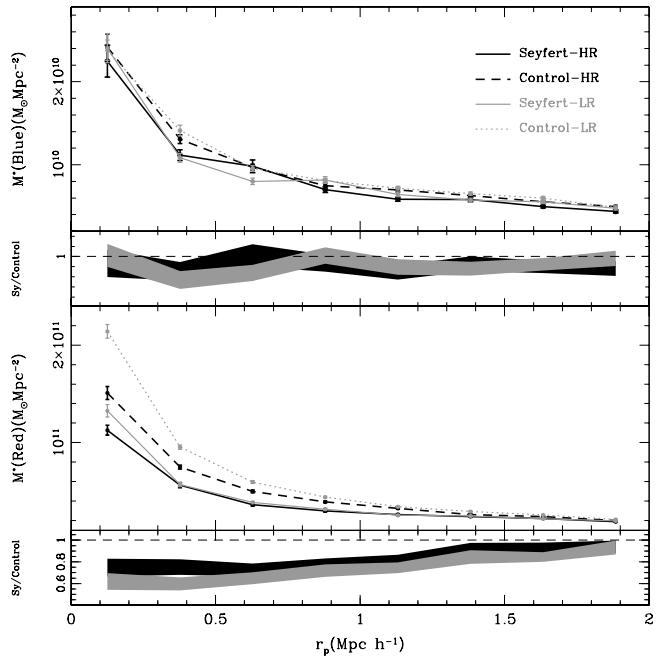}
\caption{ Differential mass profiles for blue galaxies ($\rm M_g-M_r < 0.75$) (top panel) and red galaxies 
($\rm M_g-M_r<0.75$) (bottom panel) as a function to the projected distance to the target samples.
The lower sub-panels show the stellar mass rate of blue and red neighbouring galaxies, respectively, 
between the Seyfert-HR and Control-HR 
(black shadow) samples and between the Seyfert-LR and Control-LR (grey shadow) samples. 
The shaded regions show the uncertainties computed using the bootstrap resampling technique. 
}
\label{fig:md}
\end{figure}

The particular characteristics of number density and blue fraction
around Seyfert 2 and control environments as found in the previous
section motivates us to explore any dependence in the 
stellar mass properties of galaxy neighbourhoods around Seyfert 2 and
control samples. In Fig. \ref{fig:md} the differential mass profiles
for blue (top panel) and red (bottom panel) galaxy neighbours are
shown. As is expected the red galaxy neighbourhood is significantly
more massive than the blue galaxy neighbourhood. In the blue
neighbourhood little difference is observed in the stellar mass
distributions between all sample neighbourhoods. This is more clearly
exhibited in the respective sub-panel that shows the stellar mass
ratio between the neighbourhood of Seyfert and control
samples. 

For the red neighbourhoods, the stellar mass
profiles around Seyfert 2 and control galaxies are significantly
different. Seyfert-LR to Control-LR ratios are on average
$15\%$ smaller than Seyfert-HR to Control-HR ratios (as
observed in the lower sub-panel.) Also worth mentioning is that the
red Control-LR neighbourhood is considerably more massive than the red
Control-HR neighbourhood, an effect more pronounced at $r_p \leq 1 \mpc$.

\begin{figure}
\includegraphics[width=90mm]{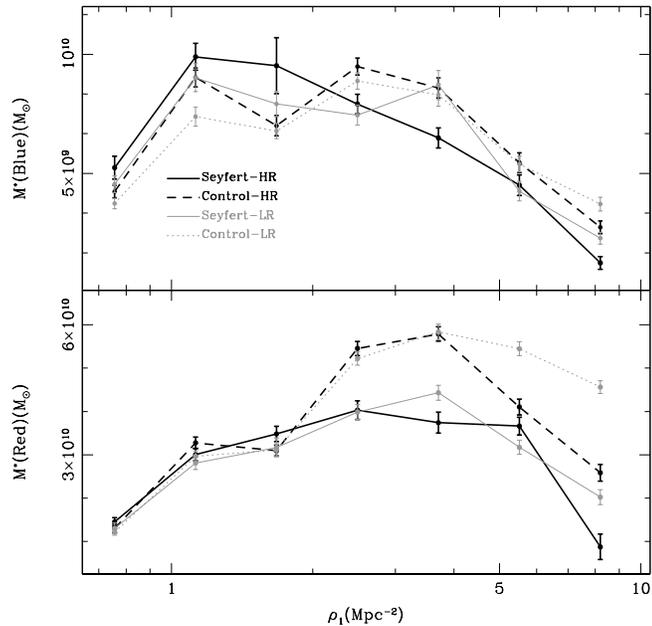}
\caption{Top: Total stellar mass (per bin) of the blue neighbouring galaxies around the target samples as 
a function of galaxy density within $1 \mpc$. Bottom:  Total stellar mass (per bin) of the red neighbouring galaxies. 
}
\label{fig:md2}
\end{figure}

We now examine if the stellar mass in galaxy neighbourhoods depends on
the galaxy number density around Seyfert 2 galaxies. In
Fig. \ref{fig:md2} we show the total stellar mass within $r_p = 1 \mpc$ 
for blue (top panel) and red (bottom panel) galaxy
neighbourhoods as a function of galaxy number density ($\rho_1$) for
each target sample. As can be seen from this figure, at low densities
($\rho_1 < 2 \rm Mpc^{-2}$) the total stellar mass of the blue galaxy
neighbours is marginally higher for the Seyfert 2 samples than for the
control samples. At higher densities a slight reversed tendency is
observed with a lack of stellar mass.

Nevertheless, in the case of red galaxy neighbours, 
the stellar masses present increasing trends up to $\rho_1 \sim 4 \rm Mpc^{-2}$
and are quite similar up to $\rho_1 \sim 2 \rm Mpc^{-2}$ around every
target sample. From $\rho_1 \sim 3 \rm Mpc^{-2}$ a stronger
distinction is observed between red neighbours of Seyfert and control
samples where the difference between Seyfert-HR and Control-HR is
approximately 2 sigma and between Seyfert-LR and Control-LR is around
6 sigmas at the highest density. The relatively small amounts of
stellar mass in galaxy neighbours (mainly for the red ones) around
Seyfert samples compared with that of control samples could suggest a
needed condition for the BH feeding where a richer gas environment 
favors the AGN activity. This is more
evident for red galaxy neighbourhood of Seyfert-LR hosts where special
conditions could generate a larger gas reservoir and also provide gas
transportation to the centre of the host galaxy.

\begin{figure}
\includegraphics[width=90mm]{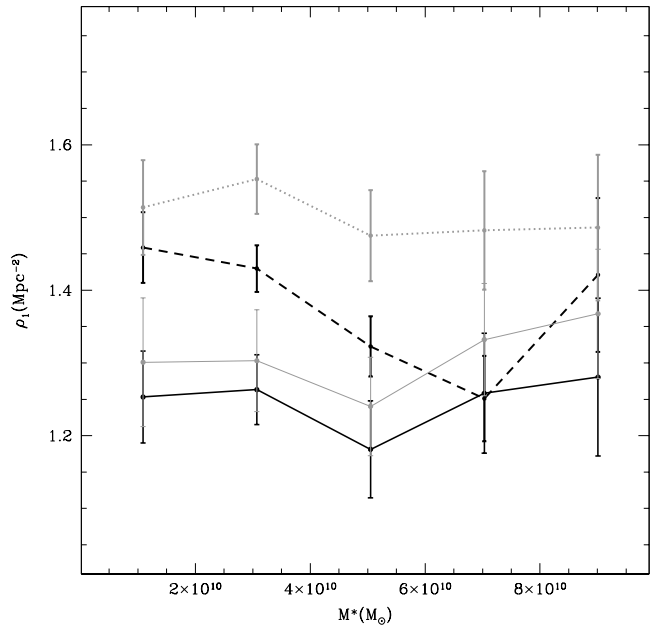}
\caption{Local number density of galaxies, within $r_p = 1 \mpc$, as a function of  
stellar mass, $M^*$, of the targets. 
}
\label{fig:md3}
\end{figure}

Finally we analysed the local number density
of galaxies around the target samples within  $r_p = 1 \mpc$c as a function of its stellar mass in Fig. \ref{fig:md3}.
A dip in the density for the control-HR galaxies of the order of the errorbars is observed in this figure.
Given the large uncertainty for this bin we are unable to distinguish if it is a
real effect or an artifact produced by the low number of objects resulting from the density calculation. The global observed tendencies of Fig. \ref{fig:md3} show that despite the similar
content of stellar mass of neighbouring galaxies, Seyfert 2 and
control samples typically live in neighbourhoods of
different density. The Seyfert 2 samples reside
in lower density environments than do their nonactive counterparts as more clearly shown in Figure 4. This result is consistent with those from \cite{monte09} who analyze a sample of red Seyfert from SDSS-DR4 and find a tendency where the AGN fraction decreases towards denser environments.

We emphasize that Seyfert-LR hosts
typically live in slightly denser environments than Seyfert-HR galaxies. 
Furthermore, the local
density in Seyfert environments does not appear sensitive to the
stellar mass of the Seyfert host.  On the other hand, the control
samples show an excess in local density for almost all bins of stellar
mass of the targets, being more significant in the Control-LR
sample. This sample having the highest galaxy number density irrespective
of target stellar mass.

\section{Discussion}

In this paper we investigate the environment of two samples of Seyfert
2 AGN, selected by the accretion rate parameter, $\cal R$. The more
rapidly growing BHs have lower-masses and are found at the centres of
younger and lower mass (stellar) galaxies than those BHs with slower
growth rates in the local universe \citep{goulding10}.  The chosen
division in $\cal R$ of the Seyfert 2 sample allowed us to examine if
environment shows a distinctive feature or characteristic on accretion
rate. This is accomplished by comparing our Seyfert-LR and Seyfert-HR
neighbourhoods with those neighbourhoods of two control samples
selected to match the distribution of redshift, luminosity,
colour, stellar mass, age and Sersic index of the Seyfert hosts.

The main results can be summarized as follows:\\

1) Even though both Seyfert 2 samples produce a large energy output
from BH accretion, both samples are found to be in different stages of
evolution as can be inferred from their host properties.  However,
when the environmental properties are considered and compared to those
of the control samples, environmental features are revealed that are
likely to be related to $\mathcal R$ selection.

2) Both Seyfert 2 samples have bluer environments than their
respective control samples, where the fraction of blue galaxies with
$M_g-M_r < 0.75$ around Seyfert-HR and Seyfert-LR are similar and
constant with projected distance to targets. However the difference of
the blue galaxy fraction between Seyfert-LR and Control-LR samples is
significantly higher than that between Seyfert-HR and Control-HR.  In
addition, the differential mass profiles of blue neighbouring galaxies
around the target samples indicate that their stellar masses are
indistinguishable. However, the stellar mass profiles of the red
galaxy neighbours show that the stellar mass of red galaxies around
Seyfert samples is lower that that around the control samples. This
effect is stronger between Seyfert-LR and Control-LR at $r_p \leq 0.5 \mpc$
reaching an excess at the 5$\sigma$ level at smaller
scales. This result suggests that the environment could compensate
any gas deficiency of the host galaxies, while at the same time
sustaining AGN activity. It should be noted that although AGN feedback 
processes can not be excluded, it is the environment of Seyfert-LR the 
most different with respect to that of  the respective control sample,
as it is evident from Fig. 4.

3) The previous results are reflected in the fact that Seyfert 2 galaxies reside in low density environments. 
There are several possibilities to explain these effects:
(i) in low density environments gas is not likely to be lost due to
stripping which can have an important impact in higher density regions
(ii) The expected higher galaxy merger rate in low density
environments allows fresh gas to be accreted; and (iii) galaxy
instabilities are also more probable in less dense environments to
provide material that feed BHs, (iv) instabilities may even be
enhanced by tidal torques that occur in encounters with low mass
galaxies.

4) The stellar mass of blue neighbours shows a small excess around
Seyfert 2 samples with respect to control galaxies at low
densities. The opposite is observed at higher densities of $\rho_{1}$
$\gtrsim$ $3$ $\rm Mpc^{-2}$ where Seyfert-HR have systematically blue
neighbours with a lower stellar mass content.  While for red
neighbours at low densities the stellar mass of Seyfert 2 and
control samples are indistinguishable, at $\rho_1 \gtrsim 3 \rm Mpc^{-2}$
the stellar mass around Seyferts is lower than for the control,
reaching a difference at the 2$\sigma$ level (Seyfert-HR vs
Control-HR), and at 6$\sigma$ (Seyfert-LR and Control-LR), at higher
densities.

5) Finally, although the low galaxy density found around Seyfert 2 within $r_p
= 1.0 \mpc$ is independent of the stellar mass of the host
galaxies and slightly higher for Seyfert-LR, in the vicinity of
galaxies of the control samples a significant difference between
Control-HR and Control-LR is found.  This would suggest than AGN do
not follow the standard morphological-segregation relation expected
for non-active galaxies.
We suggest that AGN occurrence is higher in lower density regions due to the combination of a
higher merger rate and a relative quiescent nature of the gas-dynamical processes in comparison to high density environments.

\section{Acknowledgments}

We would like to thanks to anonymous referee for the comments that helped to improve 
the paper. We thanks to Emilio Donoso for useful comments and discussion.

Funding for the SDSS and SDSS-II has been provided by the Alfred P. Sloan 
Foundation, the Participating Institutions, the National Science Foundation, 
the U.S. Department of Energy, the National Aeronautics and Space Administration, 
the Japanese Monbukagakusho, the Max Planck Society, and the Higher Education 
Funding Council for England. The SDSS Web Site is \emph{http://www.sdss.org/}.

The SDSS is managed by the Astrophysical Research Consortium for the Participating 
Institutions. The Participating Institutions are the American Museum of Natural 
History, Astrophysical Institute Potsdam, University of Basel, University of 
Cambridge, Case Western Reserve University, University of Chicago, Drexel University,
Fermilab, the Institute for Advanced Study, the Japan Participation Group, Johns 
Hopkins University, the Joint Institute for Nuclear Astrophysics, the Kavli Institute 
for Particle Astrophysics and Cosmology, the Korean Scientist Group, the Chinese 
Academy of Sciences (LAMOST), Los Alamos National Laboratory, the Max-Planck-Institute 
for Astronomy (MPIA), the Max-Planck-Institute for Astrophysics (MPA), New Mexico 
State University, Ohio State University, University of Pittsburgh, University of 
Portsmouth, Princeton University, the United States Naval Observatory, and the University 
of Washington.

{}

\label{lastpage}


\begin{thebibliography}{}

\bibitem[Abazajian \etal (2009)]{abazaja09} Abazajian K.N. \etal, 2009, \apjs, 182, 543.

\bibitem[Antonucci (1993)]{Anto93} Antonucci R., 1993, ARA\&A, 31, 473.

\bibitem[Baldwin, Phillips \& Terlevich (1981)]{BPT81}
Baldwin J. A., Phillips M. M. \& Terlevich R., 1981, \pasp, 93, 5.

\bibitem[Barrow, Bhavsar \& Sonoda (1984)]{barrow84}Barrow J.D., Bhavsar S.P. \& Sonoda B.H.,
1984, \mn, 210, 19. 

\bibitem[Begelman, Blandford \& Rees (1980)]{BBR80} Begelman M. C., 
Blandford R. D. \& Rees M. J., 1980, \nature, 287, 307.

\bibitem[Blanton \etal (2003)]{blanton03}Blanton M. R., Hogg, D. W.,
Bahcall N. A. \etal, 2003, \apj, 594, 186.

\bibitem[Blanton \etal (2005)]{blanton05}Blanton M.R., Eisenstein D., Hogg D.W., 
Schlegel D.J  \& Brinkmann J., 2005, \apj, 629, 143.

\bibitem[Brinchmann \etal (2004)]{brinch04} Brinchmann J., Charlot S.,
White S. D. M., Tremonti C., Kauffmann G., Heckman T. \& Brinkmann J., 
2004, \mn, 351, 1151. 

\bibitem[Calzetti \etal (2000)]{calzetti00} Calzetti D., Armus L., Bohlin R.C.,
Kinney A.L., Koornneef J. \& Storchi-Bergmann T., 2000, \apj, 533, 682.

\bibitem[Caon, Capaccioli \& D'Onofrio (1993)]{caon93} Caon N., Capaccioli 
M. \& D'Onofrio M, 1993, \mn, 265, 1013.

\bibitem[Capetti \& Baldi (2011)]{CapB11} Capetti A. \& Baldi R.D., 2011, \aap, 529, 126.

\bibitem[Coldwell \& Lambas (2003)]{coldwell03} Coldwell G. V. \& Lambas D. 
G.,2003, \mn, 344, 156.

\bibitem[Coldwell \& Lambas (2006)]{coldwell06} Coldwell G. V. \& Lambas D. 
G., 2006, \mn, 371, 786.

\bibitem[Coldwell \etal (2009)]{coldwell09} Coldwell G.V., Lambas D.G., S{\"o}chting
I.K. \& Gurovich S., 2009, \mn, 399, 88.

\bibitem[Connolly \& Szalay (1999)]{cono99} Connolly A.J. \& Szalay A.S., 1999, \aj, 117, 2052.

\bibitem[Dom\'inguez, Muriel \& Lambas (2001)]{dominguez01}
Dom\'inguez M., Muriel H. \& Lambas D.G., 2001, \aj, 121, 1266.

\bibitem[Dressler (1980)]{Dress80} Dressler A., 1980, \apj, 236, 351.

\bibitem[Dultzin-Hacyan \etal (1999)]{dultin99}Dultzin-Hacyan D., Krongold Y., Fuentes-Guridi I. \&
Marziani, P., 1999, \apj, 513, 111.

\bibitem[Gadotti \& Kauffmann (2009)]{GadoK09}Gadotti D.A.\& Kauffmann G., 2009, \mn, 399, 621. 

\bibitem[Gaskell (1985)]{gaskell85} Gaskell C. M., 1985, \nature, 315, 386.

\bibitem[Gonz\'alez-Mart\'in \etal (2009)]{GM09a}Gonz\'alez-Mart\'in O., Masegosa J., 
M\'arquez I. \& Guainazzi M., \apj, 2009, 704, 1570.

\bibitem[Goulding \etal (2010)]{goulding10} Goulding A. D., Alexander D. M., Lehmer B. D., \& Mullaney J.R., 2010, \mn, 406, 597.

\bibitem[Graham (2008)]{Graham08}Graham A.W., 2008, \apj, 680, 143.

\bibitem[Greene \etal (2010)]{Grene10} Greene J.E., Peng C.Y., Kim M. \etal, 2010, \apj, 721, 26.

\bibitem[G{\"u}ltekin \etal (2009)]{Gulte09}G{\"u}ltekin K., Richstone D.O., Gebhardt K. 
\etal, 2009, \apj, 698, 198.

\bibitem[Heckman \etal (2004)]{heck04} Heckman T. M., Kauffmann G., Brinchmann J., Charlot S., Tremonti C., 
White S. D. M., 2004, \apj, 613, 109.

\bibitem[Heckman \etal (2005)]{heck05} Heckman T. M., Ptak A., Hornschemeier A., Kauffmann G., 2005, 
\apj, 634, 161.

\bibitem[Hernquist \& Mihos (1995)]{HM95}
Hernquist L. \& Mihos J.C., 1995, \apj, 448, 41.

\bibitem[Hao \etal (2005)]{Ho05} Hao L., Strauss M.A., Tremonty C.A. \etal, 2005,     


\bibitem[Kauffmann \& Haehnelt (2002)]{kauff02} Kauffmann G. \& Haehnelt M. 
G., 2002, \mn, 332, 529.

\bibitem[Kauffmann \etal (2003)]{kauff03} Kauffmann G., Heckman T. M.,
Tremonti C. \etal, 2003, \mn, 346, 1055.

\bibitem[Kauffmann \etal (2003b)]{kauff03b} Kauffmann G., Heckman T. M.,
White S. D. M. \etal, 2003, \mn, 341, 33. 

\bibitem[Kauffmann \& Heckman (2009)]{KauHe09} Kauffmann G. \& Heckman T.M., 2009, \mn, 397, 135. 

\bibitem[Kewley \etal (2001)]{Kewley01} Kewley L.J, Dopita M.A., Sutherland R.S., Heisler C.A. \& Trevena J., 2001, \apj, 556, 121.

\bibitem[Kewley \etal (2006)]{Kewley06} Kewley L.J, Groves B.,  Kauffmann G. \&
Heckman T.M., 2006, \mn, 372, 961.

\bibitem[Kollatschny, Reichstein \& Zetzl (2012)]{kolla12} Kollatschny W., Reichstein A. \& Zetzl M.,
2012, \aap, 548, 37.

\bibitem[Lake, Katz \& Moore (1998)]{LKM98}
Lake G., Katz N. \& Moore B., 1998, \apj, 495, 152.

\bibitem[Li \etal (2006)]{Li06} Li C., Kauffmann G., Wang L., White S., 
Heckman T. \& Jing Y., 2006, \mn, 373, 457.

\bibitem[Lynden-Bell (1969)]{LB69} Lynden-Bell D., 1969, \nat, 223, 690. 

\bibitem[Lupton (1993)]{lupton93} Lupton R. H., 1993, 
Statistics in Theory and Practice, Princeton Univ. Press.

\bibitem[Makino \& Hut (1997)]{Makino97}Makino J. \& Hut P., 1997, \apj, 481, 83.

\bibitem[Mamon (1992)]{Mamon92}Mamon G.A, 1992, \apj, 401, 3.

\bibitem[Marconi \& Hunt (2003)]{MH03} Marconi A. \& Hunt L., 2003 \apj, 
589, L21.

\bibitem[Montero-Dorta \etal (2009)]{monte09} Montero-Dorta A.D, Crotton D.J, Yan R.
\etal, 2009, \mn, 392, 125.

\bibitem[Osterbrock \& Miller (1989)]{OM89}Osterbrock D.E. \& Miller J.S., 1989, 
Active Galactic Nuclei. Proc. IAU Symposium $N^o$ 134.

\bibitem[Padilla, Lambas \& Gonz\'alez (2010)]{padilla10} Padilla, N., Lambas, D.G.,
\& Gonz\'alez R., 2010, \mn, 409, 936.

\bibitem[Petrosian (1976)]{petro76} Petrosian V., 1976, \apj, 209, 1.

\bibitem[Popesso \& Biviano (2006)]{PB06}, 2006 
\aap, 460, L23. 

\bibitem[Purcell \etal (2011)]{Purcell11}Purcell C.W., Bullock J.S., Tollerud E.J.,
Rocha M. \& Chakrabarti S., 2011, \nature, 477, 301.

\bibitem[Rees (1984)]{Rees84} Rees M.J., 1984, \araa, 22, 471.

\bibitem[Roos (1981)]{roos81} Roos N., 1981, \aap, 104, 218.

\bibitem[Roos (1985)]{roos85} Roos N., 1985, \aap, 294, 479.

\bibitem[Sanders, Soifer \& Scoville (1988)]{sanders88} Sanders D. B., 
Soifer B. T. \& Scoville N. Z, 1988, \science, 239, 625. 

\bibitem[S\'{e}rsic (1963)]{sersic63} S\'{e}rsic J. L., 1963, BAAA, 6, 41.

\bibitem[Shlosman, Begelman \& Frank (1990)]{SBF90} Shlosman I., Begelman M.C.
\& Frank J., 1990, \nature, 345, 679.

\bibitem[Smith, Boyle \& Maddox (1995)]{SBM95}
Smith R. J., Boyle B. J. \& Maddox S. J., 1995, \mn, 277, 270.

\bibitem[S{\"o}chting, Clowes \& Campusano (2002)]{ilona02} S{\"o}chting 
I.~K., Clowes R.~G., \& Campusano L.~E.\ 2002, \mn, 331, 569. 

\bibitem[S{\"o}chting, Clowes \& Campusano (2004)]{ilona04} S{\"o}chting 
I.~K., Clowes R.~G., \& Campusano L.~E.\ 2004, \mn, 347, 1241. 

\bibitem[Sorrentino, Radovich \& Rifatto (2006)]{SRR06} Sorrentino G., Radovich M. \& Rifatto A., 
2006, \aap, 451, 809.

\bibitem[Storchi-Bergmann \etal (2001)]{Taisa01} Storchi-Bergmann T., González Delgado R.M., 
Schmitt H.R., Cid Fernandes R. \& Heckman T., 2001, \apj, 559, 147.

\bibitem[Strand, Brunner \& Myers (2008)]{Strand08} Strand N.E., Brunner R.J. \& Myers A.D, 2008,
\apj 688 180. 

\bibitem[Strauss \etal (2002)]{strauss02} Strauss M., Weinberg D. H.,
Lupton, R. H.\etal, 2002, \aj, 124, 1810.

\bibitem[Tremaine \etal (2002)]{trem02} Tremaine S., Gebhardt K., Bender R.
\etal, 2002, \apj, 574, 740.

\bibitem[Tremonti \etal (2004)]{tremonti04} Tremonti C., Heckman T. M.,
Kauffmann, G. \etal, 2004, \apj, 613, 898. 

\bibitem[Urry \& Padovani (1995)]{Urry95}Urry M.C. \& Padovani P., 1995, PASP, 803, 107.

\bibitem[Yan \& Blanton (2012)]{YanB12}Yan R. \& Blanton M.R, 2012, \apj, 747, 61.

\bibitem[Young \& Currie (1994)]{YC94} Young C. K. \& Currie M. J., 
1994,\mn, 268, 11. 

\end{thebibliography}
\end{document}